\documentclass[aps,pra,twocolumn,floatfix,superscriptaddress,showpacs]{revtex4}
\usepackage{graphicx}
\usepackage{dcolumn}
\usepackage{float}
\usepackage{bm}
\usepackage{amssymb,bezier,epsfig,amsmath}
\usepackage{epsfig}
\input rotate
\input colordvi
\begin{document}

\title{Loss of molecules in magneto-electrostatic
traps due to nonadiabatic transitions}
\author{Manuel Lara}
\altaffiliation{Institut de Physique de Rennes, UMR CNRS 6251,
Universit\'e de Rennes I, F-35042 Rennes, France}
\affiliation{JILA and Department of Physics, University of
Colorado, Boulder, CO 80309-0440}
\author{Benjamin L. Lev}  \email{benlev@uiuc.edu}
\affiliation{{Department of Physics, University of Illinois at Urbana-Champaign, 1110 W. Green St., Urbana, IL 61801}}
\author{John L. Bohn} \email{bohn@murphy.colorado.edu}
\affiliation{JILA and Department of Physics, University of
Colorado, Boulder, CO 80309-0440}

\date{\today}

\begin{abstract}
We analyze  the dynamics of a  paramagnetic, dipolar molecule in a generic
``magneto-electrostatic'' trap where both magnetic and electric fields may 
be present.  The potential energy
that governs the dynamics of the molecules is found using a reduced
molecular model that incorporates the main features of the system.
 We discuss the shape of the trapping potentials for
different field geometries, as well as the possibility of
 nonadiabatic transitions to untrapped states, i.e., the analog of Majorana transitions in a
quadrupole magnetic atomic trap.  Maximizing the lifetime of molecules in a trap is of great concern in current experiments, and we assess the effect of nonadiabatic transitions on obtainable trap lifetimes.

\end{abstract}


\maketitle

\section{Introduction}
A primary tool for the study of ultracold matter is the magnetic trap, 
which can confine paramagnetic atoms on time scales of seconds.  Using the techniques of laser cooling, atoms can be loaded into magnetic traps---such as the quadrupole trap created with anti-Helmholtz coils---at cold enough temperatures so that only modest magnetic field gradients are required to confine the atoms.  Once loaded, the lifetime in a quadrupole trap is typically limited by collisions, thermal background radiation, or Majorana transitions~\cite{Majorana32}, wherein atomic spins passing near the zero in the field minimum are not able to adiabatically follow the field direction.  To such an atom, the quantization axis is lost and it may emerge from the trap zero ``spin-flipped" to an untrapped state~\cite{MetcalfBook99}.  

The Majorana spin-flip loss mechanism can be severe enough to quench trap lifetime before forced evaporative cooling can bring the atomic system to quantum degeneracy.  Such cooling is thought to be a necessary requirement to reach the strongly correlated regimes that hold much promise for investigating novel ultracold collisions, chemistry, connections to condensed matter, and quantum information processing with particles beyond presently studied atoms, e.g., polar molecules~\cite{Doyle04}.  To prevent such loss in atomic systems, evaporative cooling is instead performed in a secondary magnetic trap of the Ioffe-Pritchard or ``baseball"~\cite{Bergeman89,MetcalfBook99} configuration which does not possess a vanishing field magnitude near the trap minimum, and thereby preserves a quantization axis for all particles confined in the trap.  Additional techniques to minimize Majorana loss include moving trap minima in time so that the atoms experience a non-zero field on average~\cite{Cornell95}, and confining in storage rings wherein the centrifugal barrier shifts the trap center away from field minima~\cite{Meijer01}).  

Large samples of molecules in their absolute rovibronic ground state have yet to be produced at temperatures below 10 mK~\cite{Doyle04}. Sample confinement therefore requires very large magnetic fields, even for the case of a quadrupole trap~\cite{Sawyer07}.  Electrostatic~\cite{Meijer05a,Rempe05} and magnetostatic~\cite{Doyle98,Doyle07} traps
have successfully been demonstrated for molecules, and inhomogeneous static electric and magnetic fields have been combined to increase trap depth~\cite{Sawyer07}.  Lifetimes have so far been limited by collisions or blackbody radiation, but once evaporative or other cooling schemes~\cite{Doyle04,Lev08} are employed, Majorana transitions will be a dominant cause of loss in all the currently employed traps for ground state molecules.  Since the large fields necessary to create, e.g., Ioffe-Pritchard traps, are generally unobtainable, we must carefully investigate the lifetime limits imposed by Majorana transitions in the currently realizable quadrupole traps and explore possible schemes to avoid such loss mechanisms for real molecular systems.

For cold molecules that possess both magnetic and electric dipole moments, novel traps formed from both inhomogeneous magnetic and electric fields may be utilized to confine molecules in ways that mitigate the unwanted Majorana transitions.   In this paper we explore the basic physics of traps for molecules in
which both magnetic and electric fields are present, which we show can mitigate Majorana loss even for simple field configurations consisting of superimposed quadrupole and homogenous fields.  Using a reduced, but analytically solvable, model of a molecule in both electric and magnetic
fields, we explore various trap geometries that can be generated from
common coil and electrode configurations.  We identify
the zone in the trapping potential where nonadiabatic, Majorana-like transitions can occur and we introduce approximate expressions to assess such loss.  More broadly, we
note that these results are of interest to the design of Stark
decelerators~\cite{Meijer99}, where slowed molecules experience rapidly changing field
configurations.  Our main focus will be on diatomic
molecules governed by Hund's coupling case (a), since in this case
the electric and magnetic dipole moments are nontrivially related.  In particular, we use the OH molecule as a representative molecular system, but we briefly address case (b) molecules as well.

\section{Molecular structure in the presence of crossed fields}

 Before discussing the Majorana spin-flip problem for representative traps and molecules, we first examine perturbations to the molecular structure in crossed electric and magnetic fields.
 We consider a diatomic, heteronuclear molecule that has a permanent
 electric dipole moment ${\bm \mu}_e$ lying along its molecular axis.  The
 molecule is moreover assumed to be paramagnetic, with magnetic
 dipole moment ${\bm \mu}_m$.  In the presence of spatially varying
 magnetic (\boldmath$\cal B$\unboldmath)  and electric  (\boldmath$\cal E$\unboldmath) fields,
 the molecule is governed by the Hamiltonian
\begin{equation} 
H = H_0-{\bm \mu}_m \cdot \mbox{\boldmath$\cal B$}({\bm r})-{\bm \mu}_e \cdot \mbox{\boldmath$\cal E$}({\bm r}), \label{Ham}
\end{equation}
where $H_0$ describes the internal workings of the molecule,
including lambda doubling, fine structure, and hyperfine
structure, if applicable.  The eigenenergies of $H$ will vary in space according the spatial variation of the fields, and these varying energies define the trap potentials.

Eigenvalues of $H$ can be constructed to any desired degree of
accuracy, generating realistic trap potentials for any desired
molecule.  Such potentials were generated for OH in Ref.~\cite{Sawyer07} according to this procedure.  Presently, however, we are interested in the general features
of the traps, and so will opt for a simplified molecular structure, albeit one which encapsulates all the relevant physics.
We will explicitly consider Hund's case (a) molecules with
$^2\Pi$ electronic symmetry (as in
OH) as well as Hund's case (b) molecules with $^2\Sigma$ symmetry.
These electronic structures are the most commonly encountered examples of open-shell diatomics, and
throughout we will neglect hyperfine structure.

\subsection{Hund's case (a) $^2 \Pi $ molecule}\label{antes}

Case (a) molecules are characterized by a strong spin-orbit
interaction that couples the electronic spin to the electronic
orbital angular momentum, and thus also to the molecular axis. The relevant quantum numbers for a case (a) molecule are $|(\Lambda
\Sigma)J M \Omega \rangle$, where $\Lambda$ and $\Sigma$ are the
projections of the orbital and spin angular momenta of
 the electrons on the molecular axis, respectively; $J$ is the total angular momentum
of the molecule, resulting from the addition of the  orbital and spin angular momentum
 of the electrons, and the rotation of the nuclei; $M$ is the projection
 of $J$ on the laboratory axis; and $\Omega$  its projection on the molecular axis.
 We will suppress $\Lambda$
and $\Sigma$ quantum numbers which take assumed values
 hereafter.

In the absence of an external electric field, the molecule is also
an eigenstate of parity, characterized by another quantum number,
$\epsilon=\pm 1$, which is conventionally denoted by $e$ and $f$,
respectively:
 \begin{eqnarray}
 | (e/f) J M \bar{\Omega} \rangle = 1/ \sqrt{2}( | J M  \bar{\Omega} \rangle + \epsilon | J M\  -\bar{\Omega} \rangle
 ),
\nonumber \end{eqnarray} where $\bar{\Omega}=|\Omega|$ is defined to
be positive and the parity of the state is given by $\epsilon
(-1)^{J-S}$.
 The energy gap between these two states due to $\Lambda$-doubling~\cite{Brown} is denoted by $\Delta$ (which equals $2\pi\times1.7$ GHz for OH).  OH molecules in their ground state are easily polarizable due to their relatively small $\Lambda$-doubling and large
electric dipole moment ($\mu_e = 1.67$ D).  In an electric field, the superposition of opposite 
parity states gives rise to an effective polarization.  

For simplicity, we will assume that  $J=1/2$ and thus $\bar{\Omega}=1/2$ (we will suppress them also in the notation), and will consider only
the four states $ |f, M=+1/2\rangle$, $|f, M=-1/2\rangle$, $ |e, M=+1/2\rangle$ and
$ |e, M=-1/2\rangle$. We will further assert that the fields applied are
small enough that $J$ remains approximately conserved.  Unfortunately,
 there exists an almost complete cancelation of
orbital and spin magnetic dipole moment for a $^2 \Pi_{1/2}$ molecule,
 leading to a nearly
vanishing Zeeman interation.  This
deficiency in our model could be easily eliminated by considering a $^2
\Pi_{3/2}$ state, but at the expense of introducing additional $M$
levels.  To maintain analytical simplicity, we will formally use a
$^2\Pi_{1/2}$ molecule,  but ascribe to it a molecular g-factor of 4/5
which is the value corresponding to OH's ground state.  This {\em ad hoc} parameter adjustment will not change the qualitative conclusions of the
model, as we have checked by comparison with the results for a
more complicated  $^2
\Pi_{3/2}$ model.

\begin{figure}
\begin{center}
\epsfig{file=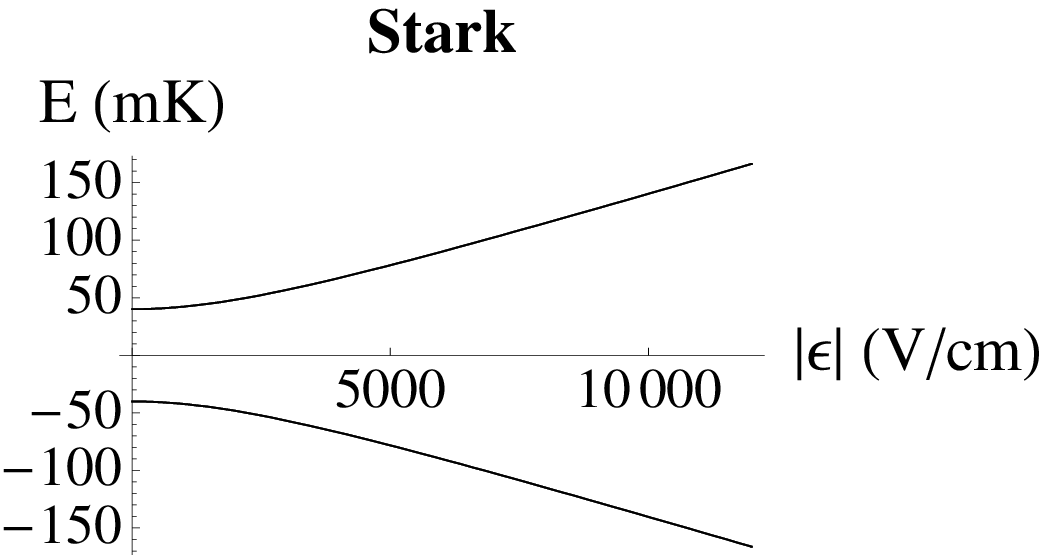,angle=0,width=0.7\linewidth,clip=}
\epsfig{file=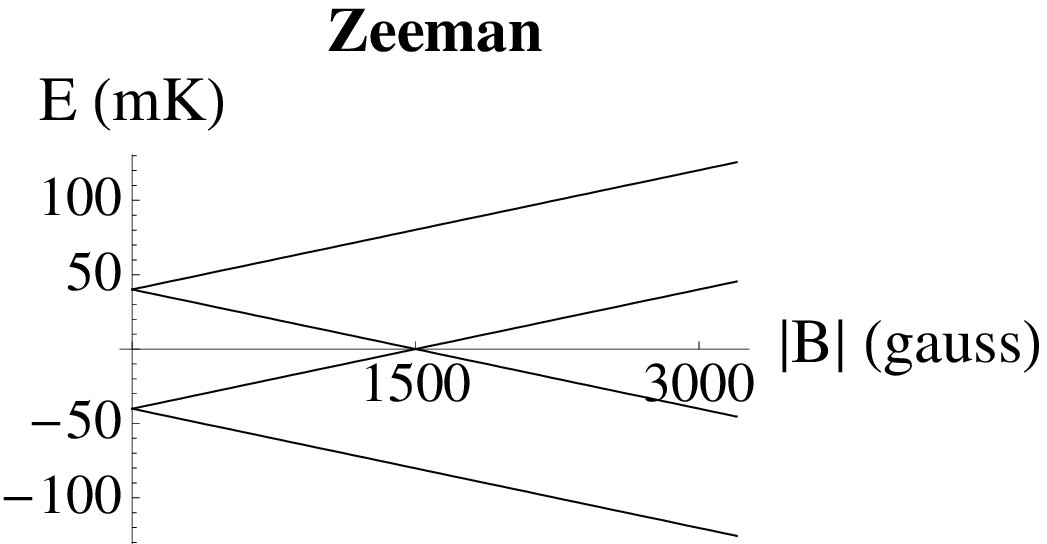,angle=0,width=0.7\linewidth,clip=}
\end{center}
\caption{ Stark (top
panel)  and Zeeman (bottom
panel) energies as a function
of the respective field magnitude for the four-state model defined
in the 
text, with parameters corresponding to a $^2\Pi_{1/2}$ OH molecule (except for a   $\sim$1 $\mu_B$ magnetic moment). The states' energies are $\pm \Delta /2$ in the absence of fields.
In the case of Stark effect, both the  weak and strong field-seeking energies are doubly-degenerate.
 The Zeeman effect breaks the degeneracy except when $U_m=\Delta /2$.  The matrix in Eq.~(\ref{Hami}) is diagonal and each line in the figure corresponds to one  $ |(e/f), M=\pm 1/2 \rangle$ basis vector.} \label{ZeeS}
\end{figure}

Using this molecular basis and without losing generality, we begin by assuming there is a magnetic field 
along the $Z$-axis at all points in space and an electric field in the $XZ$-plane
 which makes an angle $\alpha$ with respect to
the magnetic field.  The matrix elements of $H$ are easily calculated (see
Appendix~\ref{matrix elements}):
\begin{widetext}
\begin{eqnarray} H=
\left(\begin{array}{cccc} - \Delta /2 + U_m  & 0 &
 -U_e \cos \alpha  & -U_e \sin \alpha \\ 0 & - \Delta /2 - U_m   & -U_e \sin \alpha  & U_e \cos \alpha
 \\ -U_e \cos \alpha  & -U_e \sin \alpha  & \Delta /2 + U_m
 & 0  \\  -U_e \sin \alpha  & U_e \cos \alpha  &
 0 & \Delta /2 - U_m   \end{array}\right),\label{Hami}  \end{eqnarray}
\end{widetext}
where
\begin{eqnarray}
U_e &=& g_e\mu_e |\mbox{\boldmath$\cal E$}|  \nonumber \\
U_m &=& g_m\mu_B |\mbox{\boldmath$\cal B$}|, 
\end{eqnarray}
are characteristic electric and magnetic energies and $\mu_B$  is the  Bohr magneton.  Geometric factors and  any dependence on the particular 
quantum numbers of the state have been gathered 
into a global g-factor which can interpreted as the effective 
coupling with the field. Explicit expressions are given in 
Appendix~\ref{matrix elements}.  This
matrix can be diagonalized analytically at every point in space, yielding the following energy eigenvalues:
\begin{widetext}
\begin{eqnarray}
U_{1,2,3,4}= \pm \sqrt{(\Delta/2)^2 + U_e^2 + U_m^2 \pm 2 U_m
\left[(\Delta/2)^2 + (U_e \cos \alpha )^2\right]^{1/2}
 },
 \label{eigenvalues}
\end{eqnarray}
\end{widetext}
 where we have assumed  the existence of four different nondegenerate 
eigenvalues, and labeled them in increasing order of energy (i.e., ``$U_1$" labels the lowest energy eigenvalue). 
The associated eigenvectors will be denoted $|1\rangle,|2\rangle\,|3\rangle$ and $|4\rangle$. In the limit where one field vanishes, this expression
reduces to the familiar Stark and Zeeman effects (see Fig.~\ref{ZeeS}).  

\begin{figure}
\setlength{\unitlength}{4mm}
\begin{center}
\epsfig{file=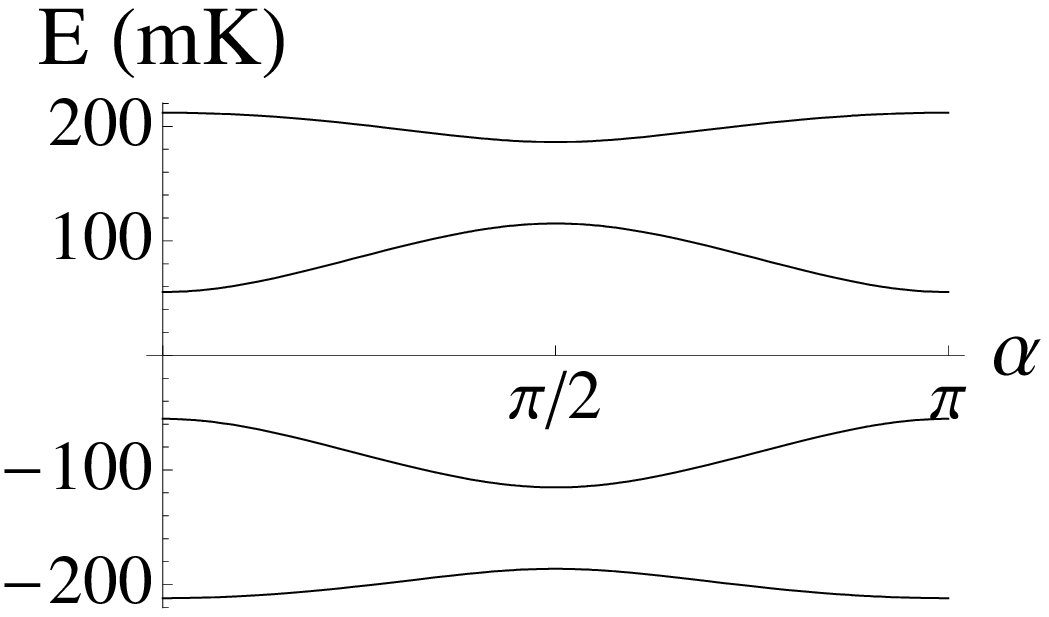,angle=0,width=0.7\linewidth,clip=}
\epsfig{file=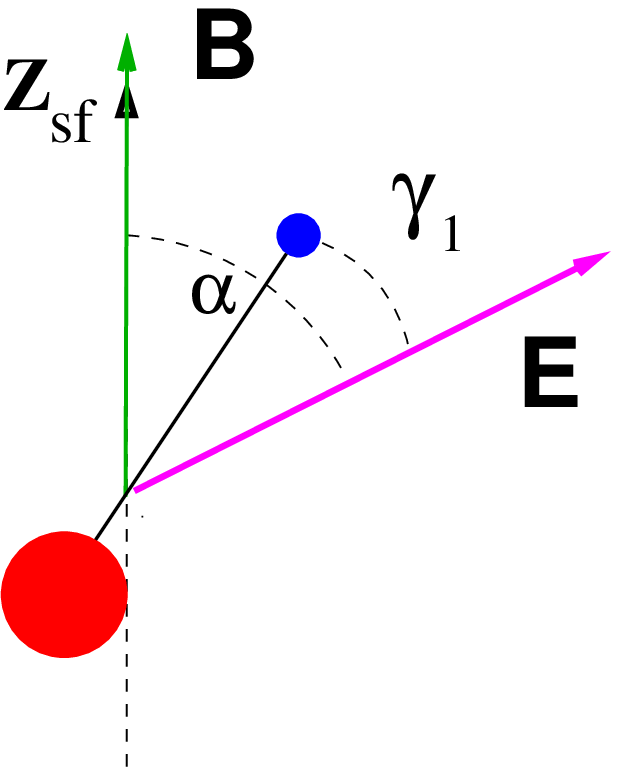,angle=0,width=0.25\linewidth,clip=}
\put(-10.3,.1){\makebox(0,0){{{ {\tiny $| 1 \rangle $ }}}}}
\put(-10.3,2){\makebox(0,0){{{ {\tiny $| 2 \rangle $ }}}}}
\put(-10.3,4.2){\makebox(0,0){{{ {\tiny $| 3 \rangle $ }}}}}
\put(-10.3,6.1){\makebox(0,0){{{ {\tiny $| 4 \rangle $ }}}}}
\caption{(Color online) Dependence of the eigenenergies of the
four-state model on the angle $\alpha$ between the electric and
 magnetic field directions for fixed values of their magnitudes (see righthand drawing).} \label{angulo}
\end{center}
\end{figure}

If the electric and magnetic fields are everywhere parallel, then
the total energy is simply the sum of the independent Stark
and Zeeman energies.  In general, however, we must account for the
angle $\alpha$ between these fields.  The eigenenergies are plotted
versus $\alpha$ in Figure \ref{angulo} for the four states.
 We assume $U_e$ and $U_m$ values that correspond to our model OH molecule
in fields of $|\mbox{\boldmath$\cal E$}|=5\times10^3$ V/cm and $|\mbox{\boldmath$\cal B$}|=5\times10^3$ G.  This figure
illustrates that the states experience avoided
crossings in the presence of non-zero $\Lambda$-doubling
 for  $\alpha = \pi/2$.  As we will see in further examples, such competition between electric and magnetic effects 
enrich the trap potentials. A semi-classical treatment of this situation is given in Appendix B.

 As an aside, we note interesting features of the energies in crossed
fields which may be useful in contexts beyond trapping. 
Applying a constant magnetic field of magnitude such that $U_m=\Delta/2$ (see Fig. \ref{ZeeS})
causes two opposite parity states to be
degenerate~\cite{Fried}, as can be seen by taking the zero-electric-field limit $U_e \rightarrow 0$ in Eq.~(\ref{eigenvalues})). Because of this degeneracy, the Stark effect is immediately linear in a nonzero electric field, regardless of the angle between fields. The slope of this linear dependence, however, may depend on $\alpha$. This may be useful for increasing the efficiency of Stark decelerators~\cite{Sawyer07b}.

Conversely, a Zeeman effect that is quadratic in $|\mbox{\boldmath$\cal B$}|$  can be obtained by applying a constant perpendicular electric field which induces an avoided crossing between two states.  Rearranging terms in
Eq.~(\ref{eigenvalues}) and substituting $\alpha=\pi/2$ into the eigenvalues, one can obtain   $U_{3,2} = \pm \sqrt{U_{e}^{2} +
(U_{m} \mp \Delta/2)^2}$. The constant electric field has induced a
quadratic Zeeman effect in the vicinity of $U_{m}=\Delta/2$.
This effect could be used to guide molecules moving in two dimensions:
in a region where a linearly varying magnetic field exists, 
a constant perpendicular electric field, for instance, induces minima at
points where the magnetic field has the value  $|\mbox{\boldmath$\cal B$}| =\Delta /2 \mu_B g_m $ ($\sim$1.5 kG  for our model OH).

\subsection{Hund's case (b) molecules}
Molecules like CaH, the first molecular system trapped by the buffer gas cooling method~\cite{Doyle98}, or
YbF, whose trapping could be useful to measure the electric dipole
moment of the
 electron~\cite{Hudson02}, belong to the category of molecules known as Hund's case (b)~\cite{Brown}.
We consider here a pure case (b) molecule of symmetry $^2\Sigma$, in
which case the electronic spin, hence the magnetic dipole moment,
is completely decoupled from the molecular axis. Thus,
the electric and magnetic dipoles can independently follow the
external electric and magnetic fields and the net energy is the sum
of the two energies. The behavior of such a molecule in a
magneto-electrostatic trap does not require special attention
and will not exhibit the richness of behavior enjoyed by case (a)
molecules.

For real molecules, the dipoles are only approximately independent.  The spin and rotational degrees of freedom are coupled by a Hamiltonian of the form $\gamma \bm{N} \cdot \bm{S}$, where the electronic
orbital angular momentum and the rotational angular momentum of the
nuclei couple to form $N$, with projection  $\lambda$ on the
internuclear axis ($0$ in this case).  $N$ couples to the electronic
spin $S$  to give total angular momentum $J$.   In order to
estimate its effect we can introduce the spin-rotation interaction
as a perturbation.  We have again built a simple analytical model including  only two rotational states $N=0,1$. In this case, the electric field  is the one which lies along the $Z$ axis, and we can restrict ourselves to the $M_N=0$ subset. We assume that $\gamma$, $U_e$, and $U_m$ are all much
smaller than the rotational spacing $B$.  With this perturbation, the energy of the ground weak field-seeking state varies as
$\propto \gamma^2 U_e^2 U_m \sin^2(\alpha)/B^4$ with the angle
$\alpha$.

\section{Conditions for nonadiabatic transitions to untrapped states}

We turn now to the nonadiabatic changes of internal state that lead to trap
loss as a molecule moves through a trap minimum.  Assuming an adiabatic separation in the spirit of the Born-Oppenheimer
approximation, we can distinguish the ``slow''  center of
mass coordinates of the molecule, $\bm{r}$, from the ``fast'' internal
coordinates  (denoted collectively as $q$).  The 
center-of-mass coordinates can  be treated classically, since
typical translational energies are much bigger than the trap energy
level spacing.  Thus, the molecules are described by position and
momentum coordinates $\left[\bm{r}(t), \bm{p}(t)\right]$.  Adiabatic energy
states are then  defined parametrically at each $\bm{r}$:
\begin{eqnarray}
H(\bm{r})
|\phi_k(q;\bm{r})\rangle=E_k(\bm{r})|\phi_k(q;\bm{r})\rangle,\nonumber
\end{eqnarray}
where $E_k=E_k(\bm{r})$ are the local energies and
$|\phi_k(q)\rangle=|\phi_k(q;\bm{r})\rangle$ the corresponding eigenstates.

More generally, the internal state of the molecule is a superposition of
these adiabatic  states:
\begin{eqnarray}
|\Psi(q,t) \rangle = \sum_k  a_k(t) e^{-i \omega_k(t)} |\phi_k(q;\bm{r}(t)) \rangle, \nonumber
\end{eqnarray}
 with $\omega_k(t)=\int dt' E_k(t')/ \hbar$.
By substituting the expansion of $|\Psi(q,t)\rangle$ into the
 time-dependent
Schr{\"o}dinger equation and solving for the coefficients, we find
 a system of coupled differential equations 
for the  coefficients
 $a_k$,
\begin{eqnarray}
\frac{da_j}{dt} = - \sum_k a_k \langle\phi_j|\frac{d}{dt}|\phi_k\rangle  e^{-i (\omega_k(t)-\omega_j(t))}\label{adia}.
\end{eqnarray}
From normalization constraints, $\langle\phi_j|\frac{d}{dt}|\phi_j\rangle=0$, $\delta_{jk}$ is the Kronecker delta function, and $\hbar$ is Planck's constant.

We are concerned with transitions from ``trapped'' states
$|\phi_t\rangle$, which possess potential minima in the center of the trap, to
``untrapped'' states $|\phi_u\rangle$, with maxima that push the molecules far away from the trap center. 
  The probability $p_{t \rightarrow u }$ of undergoing a nonadiabatic transition
  during a process can be found to be at most the order of~\cite{Mess}
\begin{eqnarray}
p_{t \rightarrow u } \lesssim max
\left| \frac{\langle\phi_t|\frac{d}{dt}|\phi_u\rangle}{|E_t-
  E_u|/\hbar} \right| ^2, \label{salto2}
 \end{eqnarray}
  and accordingly, adiabaticity  requires
\begin{eqnarray}
|\langle\phi_t|\frac{d}{dt}|\phi_u\rangle|\ll\frac{|E_t- E_u|}{\hbar}. \label{salto}
 \end{eqnarray}

To avoid calculating time derivatives of the states---usually a difficult task---we can recast the matrix elements using the  Hellmann-Feynman theorem~\cite{Hellmann37,Feynman39},
 \begin{eqnarray}
\langle\phi_t|\frac{d}{dt}|\phi_u\rangle=-\frac{\langle\phi_t|(\frac{d H}{dt})|\phi_u\rangle}{\hbar(E_t- E_u)}.\nonumber
 \end{eqnarray}
Furthermore, the time derivative can be re-expressed in a more intuitive
way using the velocity $\bm{v}= \frac{d \bm{r} }{dt}$
 of the molecule and $\frac{d}{dt}= {\nabla}_{\bm{r}}  \cdot \bm{v}$.
The adiabaticity criterion then becomes
\begin{equation}
|\langle\phi_t|({\nabla}_{\bm{r} } H)|\phi_u\rangle \cdot \bm{v}|\ll\frac{(E_t- E_u)^2}{\hbar}.\label{prob}
 \end{equation}
Evaluating $\nabla H_r$ is relatively simple once the distribution of fields is specified.
Distinguishing between the different radial, polar, and azimuthal spherical
components of the velocity will be useful, and we can use the  expansion:
 $\bm{v}= v_r \hat{r}+v_{\theta} \hat{\theta}+v_{\phi} \hat{\phi}$, with $\hat{r}$, $\hat{\theta} $, and $\hat{\phi} $ 
the unit vectors in radial, polar, and azimuthal directions, respectively.  

As initial examples, we first examine the limiting cases of traps with either $U_e=0$ or $U_m=0$ to gain intuition before analyzing traps formed from crossed electric and magnetic fields.

\subsection{Magnetostatic traps}\label{magtraps}

In the case of the magnetic quadrupole trap, Eq.~(\ref{prob}) reduces to the results well-known from previous treatments~\cite{MetcalfBook99}.  We demonstrate this assuming a magnetic field
distribution $\mbox{\boldmath$\cal B$}(\bm{r})=\nabla {\cal B}(x,y,-2z)$, where $\nabla {\cal B}$ is the magnetic field gradient on the $XY$ plane.  We first consider the situation in which particle's trajectory is restricted to the $XY$-plane.  Nonadiabatic transitions can connect the weak field-seeking
state $|\phi_t\rangle=|f,M=+1/2\rangle$ to the strong field-seeking state with the same
parity $|\phi_u\rangle=|f,M=-1/2\rangle$.  (The first quantum number is the parity, and $M$ denotes the projection on the local field axis throughout the remainder of this paper.)  This transition probability is given by
 \begin{eqnarray}
|\langle\phi_t|\frac{d}{dt}|\phi_u\rangle|=\frac{v_{\phi}}{2r}, \label{mativ}
 \end{eqnarray}
and the energy gap is $|E_t- E_u|=2 U_m$. As the magnetic field in
this plane is given by
 $|\mbox{\boldmath$\cal B$}(\bm{r})|=\nabla {\cal B}r$, Eq.~(\ref{salto})
establishes the adiabaticity criterion
 \begin{eqnarray}
\frac{v_{\phi}}{2r}  \ll \frac{2 \nabla {\cal B} r \mu_B g_m}{\hbar},\label{kkl}
 \end{eqnarray}
for an azimuthal velocity $v_{\phi}$ at a distance $r$ from the
center.

One can define a ``zone of death" within which
the molecule will be lost with high probability due to nonadiabatic
transitions.  We can roughly determine 
 its size using the radius at which
 Eq.~(\ref{kkl}) becomes an equality: 
 \begin{eqnarray}
r_{\phi}  \approx ( \hbar v_{\phi} / 4 g_m \mu_B \nabla {\cal B})^{1/2}. \label{quadrum}
 \end{eqnarray}
 The subscript $\phi$ is a reminder that these transitions occur due to the molecules' motion in the ${\hat \phi}$ direction, orthogonal to the local magnetic field.
 Spin-flips to untrapped states (Majorana transitions) are thus expected within a zone close
to the center of the quadrupole distribution defined by $r \lesssim r_{\phi} $.  In this zone, 
 the change in
the orientation of the field (and thus in the adiabatic basis) is rapid when the molecule is on a trajectory close to the field zero, and the energy gap between trapped and untrapped states becomes small. Both circumstances being essential in Eq.~(\ref{salto}), a quick change in the internal state may happen. Motion in the radial direction does
not cause any transitions because the orientation of the field does
not change along this direction  ~\footnote{This does not hold true if the molecule traverses the exact field minimum.   In such a case, which is of low probability, a spin-flip relative to the local direction of the magnetic field occurs due to the fast 180$^{\circ}$ inversion of the field direction.}.   Thus, the size of the zone depends on the azimuthal component of the velocity. 
  The estimate given by Eq.~(\ref{quadrum}) can also be
obtained intuitively, by requiring  the Larmor precession frequency
(proportional to the energy gap) be less than the angular
velocity (proportional to the rate of change of field orientation), which ensures that the dipole is able to adapt to the local field
\cite{Cornell95,Ketterle96}. 

For molecules moving in the $XZ$- or $YZ$-planes, the radius of the nonadiabatic zone shows the anisotropy 
induced by the different gradient along the $Z$ axis.  In general, the  dependence   on the polar position is given by 
\begin{eqnarray}
r_{\theta}(\theta) \approx \left( \frac{\hbar v_{\theta} }{ 2 g_m \mu_B
\nabla {\cal B} (1+3 \cos^2 \theta)^{3/2}} \right)^{1/2}. \label{XZ}
 \end{eqnarray}
 Moving across the  $X$- and $Z$-axes, this reduces to
 \begin{eqnarray}
r^X_{\theta} \approx (\hbar v_{\theta} / 2 g_m \mu_B \nabla {\cal B})^{1/2},
 \end{eqnarray}
 and
\begin{eqnarray}
r^Z_{\theta}  \approx ( \hbar v_{\theta} / 16 g_m \mu_B \nabla {\cal B})^{1/2}\label{enXZ}.
 \end{eqnarray}
The former differs from Eq.~(\ref{quadrum}) due to the component of the velocity $v_{\theta}$ considered.  The
rate of change of the orientation of the field that the molecule experiences naturally depends 
on the direction of movement:  the change is slower when moving 
 in the azimuthal direction (on the $XY$ plane, Eq.~(\ref{quadrum})) than 
in the polar direction, (on the $XZ$ plane, Eq.~(\ref{enXZ})).  In summary, a different 
radius for the zone of death can be assigned to every direction,
 to every nonadiabatic transition from trapped to non-trapped state, and to every component of the velocity.  For the sake of making order-of-magnitude estimates of trap loss, however, one may simply use Eq.~(\ref{quadrum}) for every position and direction~\cite{Cornell95}. Finally, it is worth noting that Brink {\em et al.}~\cite{Brin} have recently calculated loss rates for atoms with hyperfine structure trapped in typical magnetic configurations.

\subsection{Electrostatic traps}\label{Sec}

For a quadrupole electric field distribution, 
$\mbox{\boldmath$\cal E$}(\bm{r})=\nabla {\cal E}(x,y,-2z)$, with $\nabla {\cal E}$ the field gradient on the $XY$ plane, nonadiabatic
transitions require more elaboration.  There exist only two different eigenvalues and given the degeneracy we have to define a particular adiabatic basis. We will use the $M$ projection on the local field to label the vectors within each doubly-degenerate subspace, adding a subindex $1$ or $2$ to distinguish $M=+1/2$ from $M=-1/2$, respectively.
  Specifically, the molecules are trapped in the electric weak-field-seeking states $|\phi_{t1}\rangle$ and $|\phi_{t2}\rangle$ which  can make nonadiabatic transitions to the untrapped
states $ |\phi_{u1}\rangle$ and $ |\phi_{u2}\rangle$, which are also degenerate.  The latter are lower in energy by $\Delta$ from the trapped states at zero field (see Fig.~\ref{ZeeS}).

Using the formalism presented above, we identify 
two types of transitions, spin conserving and spin changing, which turn out to depend on different components of velocity: 
\begin{eqnarray}
|\langle \phi_{t1}|\frac{d}{dt}|\phi_{u1}\rangle|=|\langle \phi_{t2}|\frac{d}{dt}|\phi_{u2}\rangle| & = & \label{u1} \\ 
\frac{(\Delta/2)  g_e \mu_e  v_{r} \nabla {\cal E}}{2[(\Delta/2)^2 + U_e^2]} & \ll &  \frac{2 \sqrt{(\Delta/2)^2 +
U_e^2}}{\hbar}  \nonumber\\ 
|\langle \phi_{t1}|\frac{d}{dt}|\phi_{u2}\rangle|=|\langle \phi_{t2}|\frac{d}{dt}|\phi_{u1}\rangle| & = &  \label{u2} \\ 
\frac{ g_e \mu_e  v_{\phi}\nabla {\cal E} }{2\sqrt{(\Delta/2)^2 + U_e^2} }  & \ll &  \frac{2 \sqrt{(\Delta/2)^2 +
U_e^2}}{\hbar}. \nonumber
\end{eqnarray}
Note that they coincide when evaluated at the center of the trap, where the symmetry blurs the difference between radial and azimuthal velocities.
 Solving for the radius of the zone of nonadiabatic transitions yields for the $M$-changing transitions:
 \begin{equation}
r_{\phi} \approx  \frac{\left( \hbar g_e   \mu_e v_{\phi} \nabla {\cal E}  / 4   -
(\Delta/2)^2 \right)^{1/2} }{g_e \mu_e \nabla {\cal E} }\label{radio1}.
 \end{equation}
 This process is completely analogous to the change of $M$ in the
magnetic field case, modified to accommodate the 
$\Lambda$-doubling.  Indeed, if $\Delta=0$, then Eq.~(\ref{radio1}) is
identical to that for the magnetic case, with magnetic
quantities replaced by electric ones. Motion in the azimuthal direction must be slow enough for the molecule to be able to change its dipole
{\em orientation} with the local electric field. Otherwise, it
will undergo a nonadiabatic transition.  

Transitions which change only the parity composition but preserve $M$ depend on the
radial velocity $v_r$.  This criterion leads to a slightly different radius for nonadiabatic transitions than that for the $M$-changing transitions:
 \begin{eqnarray}
r_{r} \approx \frac{ \left[ (  \hbar g_e   \mu_e v_r \nabla {\cal E}  \Delta / 8  )^{2/3}-(\Delta /2)^2 \right]^{1/2} }{g_e \mu_e\nabla {\cal E} }.\label{radio2}
 \end{eqnarray}
This represents a new kind of transition, foreign to the usual
Majorana loss mechanism involving spins in a magnetic field.  For motion in the radial
direction, $v_{r}$
must be slow enough for the molecule to change its 
{\em polarization} adiabatically with the local electric field. In other words,
the mixing of parity states $f$ and $e$ 
that compose the energy eigenstate must evolve because a molecule undergoing radial motion experiences a change in the
magnitude of the electric field generating the quadrupole trap. 

Notice that the $\Lambda$-doubling plays a vital role in the ability of
the molecule to undergo a nonadiabatic transition.  Without lambda
doubling, the eigenstates would be states of good $\Omega$ for all
field strengths, insuring the preservation of these states as the molecule moves
radially.  Moreover, nonzero $\Lambda$-doubling smoothes the potentials
near the avoided crossing, making nonadiabatic transitions less likely.
Indeed, for $v_{\phi}<\Delta^2/\hbar g_e \mu_e \nabla {\cal E}$ or, equivalently, a value of $\Delta$ larger than
$\sqrt{v_{\phi} \hbar g_e \mu_e \nabla {\cal E}}$, nonadiabatic
transitions become very unlikely. A similar expression holds for $v_{r}$.

To highlight the difference between nonadiabatic transitions in magnetic versus electric quadrupole fields, we consider the OH radical as discussed in Ref.~\cite{Sawyer07}.
Using typical field strengths and molecular velocities of 10 m/s, and
substituting relevant OH ($^2 \Pi_{3/2}$) energies in our $J=1/2$
model, the radius of the zone of nonadiabaticity would be $\sim$1 $\mu$m for a magnetic
quadrupole trap.   However, such a zone does not exist at all for the
electric quadrupole trap. Although not necessarily the case for all
molecules, zones of death for OH in an electrostatic trap seem to be
essentially negligible at the slow velocities obtainable at the terminus of a Stark decelerator.  The $\Lambda$-doubling effect eliminates this loss channel.

Loss from the electrostatic trap can also be discussed for NH in the
metastable state $^1 \Delta$. In this case, the $\Lambda$-doublet is significantly
smaller than in OH, leading to a larger zone of nonadiabatic transitions.  We estimate that this zone is on the order of 1 $\mu$m  in diameter for an electric field gradient of  $3 \times 10^{4}$ V/cm$^{2}$,    which is typical for the types of experiments described in Ref.~\cite{Meijer07b}.   The corresponding loss rate can then be
estimated as above, yielding a trap lifetime on the  order of $10^4$ s
at a temperature of 100 mK. This value is much larger than the
estimated lifetime corresponding to blackbody
radiation~\cite{Meijer07c}, and also consistent with the lifetimes found
experimentally~\cite{Meijer07b}.   However, the lifetime decreases to 1 s when the temperature is 1 mK, and nonadiabatic loss processes will become comparable to others upon cooling to temperatures of interest.

\section{Traps with crossed $\mbox{\boldmath$\cal E$}$ and $\mbox{\boldmath$\cal B$}$ fields}

Upon combining $\mbox{\boldmath$\cal E$}$ and $\mbox{\boldmath$\cal B$}$ fields, both the physics of nonadiabatic transitions
and its interpretation become more complicated.  Because the fields
are non-parallel in general, few quantum numbers are conserved, and
transitions may occur from the upper trapped state to any of the
 three others in our model.  
We examine two 
easily-realized experimental configurations.  Many more are possible, but we start here with the simplest, non-trivial examples.  Namely, we will consider the influence of a uniform electric field on a magnetic quadrupole
trap as well as the influence of a uniform magnetic field on an
electric quadrupole trap.  Nonadiabatic transition rates for these situations are found in various limits and loss rates estimated.

\subsection{Magnetic quadrupole field plus constant electric field}

We first explore the consequences of crossed fields for a
magnetic quadrupole trap with a constant electric field, as was the
situation in the recent JILA experiment described in Ref.~\cite{Sawyer07}.

\subsubsection{Trap potential}

In the case of an azimuthally symmetric quadrupole magnetic trap, the field
distribution near the trap's center is given by $\mbox{\boldmath$\cal B$}(\bm{r})=\nabla {\cal B}(x,y,-2z)$.  For a simple spin-1/2 particle (i.e.,
ignoring lambda doubling),  this results in two surfaces, one
for weak-field-seeking states and one for strong-field-seeking
states, which touch only in the center where $|\mbox{\boldmath$\cal B$}|=0$.  It is around this region that Majorana transitions are expected, as discussed in Section~\ref{magtraps}. In the
case of a molecule with a lambda doublet, however, there are two
such sets of surfaces, offset by $\Delta$ from one another. They can
be visualized by rotating the four energy curves on the lower panel of Fig.~ 
\ref{ZeeS} about its vertical axis. As can be also concluded from the figure, the high-field-seeking 
surface of $f$ parity and the low-field-seeking surface of $e$ parity  become degenerate for a particular value of the magnetic field magnitude.

\begin{figure}[t]
\setlength{\unitlength}{8mm}
\begin{center}
\includegraphics[width=8.6cm]{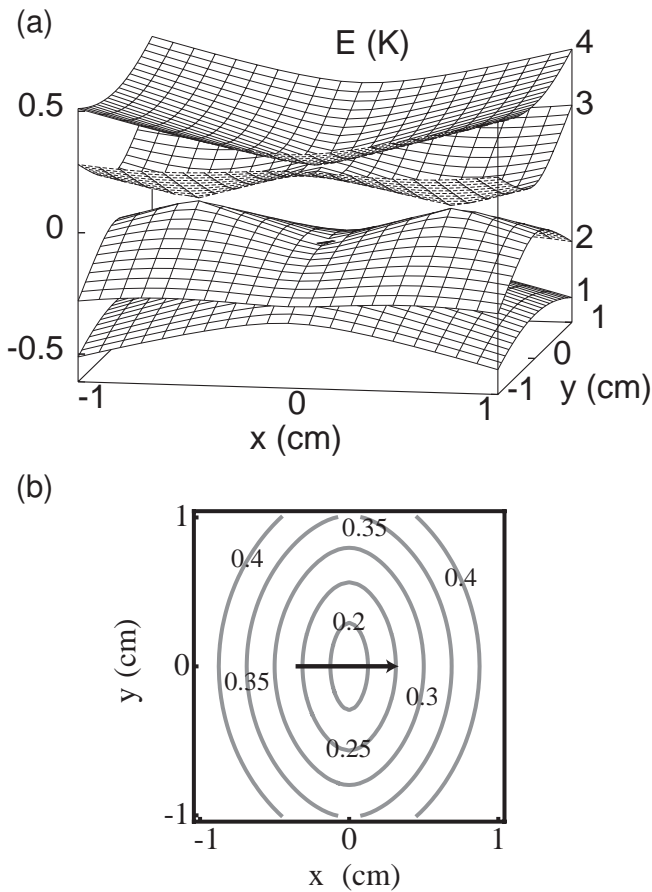}
\caption{ (Color online) a) Adiabatic trap potentials for a magneto-electrostatic
trap in which a constant electric field is superimposed on to a 
magnetic quadrupole field. Slices through the center of the 
trap ($z=0$) are shown. They correspond to the states $|1\rangle$, $|2\rangle$, $|3\rangle$, and $|4\rangle$ which are defined in the text. 
These eigenenergies have been calculated using the four-state model
(see text) for a molecule with $\Delta \approx 80$ mK, $\mu_e=1$ D, and $\mu_m\sim1$ $\mu_B$ (like OH), within
 a field distribution with parameters given  by
${\nabla \cal B}=10^4 $ G/cm, $|\mbox{\boldmath$\cal E$}_0|=1.2\times10^4$ V/cm.  Panel b) shows contours of the highest surface $|4\rangle$ around the trap center. The arrow points parallel to the electric field.} \label{constantel}
\end{center}
\end{figure}

Introducing a constant electric field $\mbox{\boldmath$\cal E$}_0 =
(|\mbox{\boldmath$\cal E$}_0|,0,0)$ to the trap shifts the energies away from
the simple, purely magnetostatic picture. Figure~\ref{constantel}(a) shows the four adiabatic surfaces
for our model OH molecule for ${\nabla \cal B}=10^4 $ G/cm and $|\mbox{\boldmath$\cal E$}_0|=1.2\times10^4$ V/cm, with eigenenergies given by Eq.~(\ref{eigenvalues}). They are denoted by $|1\rangle$, $|2\rangle$, $|3\rangle$, and $|4\rangle$, using the name of the corresponding eigenvectors.
These surfaces are depicted
as slices through the center of the trap where $z=0$.  The upper two surfaces and the lower two
remain degenerate
 at $(x,y)=(0,0)$, with energies given by the Stark eigenvalues 
\begin{equation}
\pm E_S=\pm\sqrt{(\Delta/2)^2+U_{e0}^2},
\end{equation}
 where $U_{e0}= g_e \mu_e |\mbox{\boldmath$\cal E$}_0|$ (a subindex $``0"$ is added to emphasize that the electric field is constant in space). The electric
field breaks the cylindrical symmetry of the pure magnetic
quadrupole trap.  Thus, the highest energy surface, $|4\rangle$, possesses roughly elliptical
contours of energy (Figure~\ref{constantel}(b)). Note that the trap is not harmonic, but linear in the center.

The second-highest energy surface, $|3\rangle$, which is also able to trap molecules,
has a double-well structure due to two conical crossings 
with the next surface,  $|2\rangle$, and 
which occur at the coordinates $x=\pm
 E_S /( g_m \mu_B {\nabla \cal B})$.
 At these particular points, 
electric and magnetic fields are collinear and the surfaces can
cross.  The
 crossing is avoided for any other angular configuration, which leads to the conical
structure.  Thus, while an
interesting trap geometry may be generated, 
nonadiabatic transitions may be possible at these type of points.   However, we only discuss loss in the central region of the magnetic quadrupole trap.

\subsubsection{Losses}

\begin{figure}[t]
\begin{center}
\includegraphics[width=8.6cm]{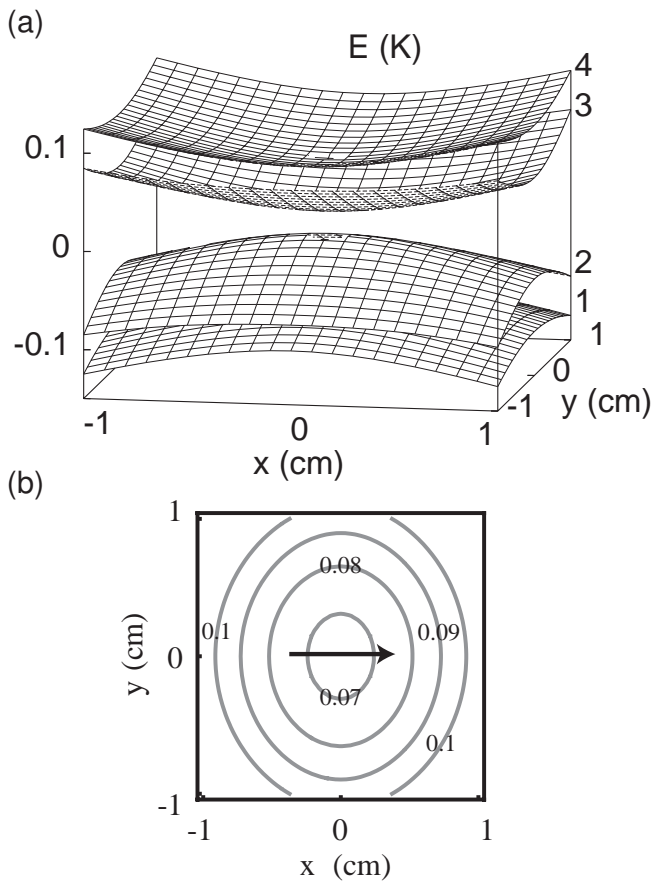}

\caption{(Color online) a) Adiabatic trap potentials for a magneto-electrostatic
trap in which a constant magnetic field is superimposed on to a 
electric quadrupole field. Slices through the center of the 
trap ($z=0$) are shown. They correspond to the states $|1\rangle$, $|2\rangle$, $|3\rangle$, and $|4\rangle$ which are defined in the text. 
These eigenenergies have been calculated using the four states model
(see text) for a molecule with $\Delta \approx 80$ mK, $\mu_e=1$ D, and $\mu_m\sim1$ $\mu_B$ (like OH), within
 a field distribution with parameters given by
${\nabla \cal E}=5\times10^3$ V/cm$^2$, $|\mbox{\boldmath$\cal B$}_0|=10^3$ G.  Panel b) shows contours of the highest surface $|4\rangle$ around the trap center. The arrow points parallel to the magnetic field} \label{constantma}
\end{center}
\end{figure}

 As a result of adding the constant electric field to the magnetic quadrupole field, nonadiabatic transitions may occur between $|4\rangle$ and $|3\rangle$ as in the purely magnetic case, but now transitions from $|4\rangle\rightarrow|2\rangle$ and $|4\rangle\rightarrow|1\rangle$ are possible due to the electric field.  

We begin with $|4\rangle \rightarrow |1\rangle$. Using $E_1 = -E_4$, the condition for adiabaticity when moving
on the $XY$-plane (with the electric field along the $X$-axis)
is:\begin{widetext}
\begin{eqnarray}
g_m \mu_B \nabla {\cal B} U_{e0}
 \left[ v_{\phi} \cos \phi  
\left(1+\frac{U_m}
{ \sqrt{(\Delta/2)^2 + U_{e0}^2\cos^2 \phi) 
}} \right) + v_r \sin \phi \right]  \ll 4  \frac{E_4^{3}}{ \hbar}  \label{exactaconsE}
\end{eqnarray}
\end{widetext}
This criterion is always satisfied in the limit
$U_{e0}\rightarrow0$, which reduces to the pure magnetic quadrupole field
 case where there is no such transition.
To gain insight from this expression, we can  assume
that the effects from the magnetic field
are small near the center of the trap ($U_m \approx 0$).  This
results in the simple  expression
\begin{eqnarray}
\hbar g_m \mu_B \nabla {\cal B} U_{e0}
 \left( v_{\phi} \cos \phi  
+   v_r \sin\phi 
  \right)  \ll 4  E_S^{3}. \label{aproximo}
\end{eqnarray}
 The combination $(  v_{\phi} \cos \phi  
+   v_r \sin\phi )$ can be easily identified as the component ($v_y$) of the velocity
that is perpendicular to the electric field, which is along $X$. Thus,
Eq.~(\ref{aproximo}) can be  re-expressed
as
\begin{eqnarray}
 \frac{ \hbar  g_m \mu_B \nabla {\cal B} v_y U_{e0}  }{ 4 \sqrt{(\Delta/2)^2+U_{e0}^2}} \ll 1, \label{electrico1}
\end{eqnarray}
which may  be numerically evaluated in a particular trap to check if
this transition leads
to losses. In fact, the maximum of the left-hand side of the previous expression as a function of $U_{e0}$ is given by
$ (\hbar  g_m \mu_B \nabla {\cal B} v_y)/(3^{3/2} \Delta^2)$, which turns out to be neglectable for realistic traps containing OH.

Analytical expressions analogous to Eq.~(\ref{exactaconsE})
can be obtained for the other two transitions, but their exact form is complicated and we do not present them here.  Instead, since  these transitions are
expected to occur near the center of the trap, we can assume that
the magnetic field interaction there is weaker than the electric
field interaction. We thus consider the magnetic field as a perturbation.
 The degeneracy of the eigenvalues
makes it convenient to use Brillouin-Wigner perturbation theory
\cite{Brillouin32,Wigner35} and a zero-order basis which diagonalizes the
Zeeman perturbation on each parity subset separately. 
For example, Eq.~(\ref{aproximo}) coincides with  the expression obtained 
using perturbation theory when keeping only leading terms for each
member of Eq.~(\ref{prob}).  For the perturbative series to converge, the condition $U_m \ll E_S$ has to be
fulfilled.
 
Using the perturbation approach to estimate the transition rate for
 $|4\rangle \rightarrow |2\rangle$, which is  azimuthal, we find it is  
 less likely to occur than the transitions $|4\rangle \rightarrow |1\rangle$ because
 the transition probability is of higher order while the
energy gap remains nearly the same.  Therefore, condition~\ref{electrico1} should set a bound on these losses as well.

The remaining transition,
 $|4\rangle \rightarrow |3\rangle$, is the only one
possible in the absence of an electric field (see Eq.~(\ref{quadrum})). 
Perturbation theory shows that the previously circular zone of nonadiablatic transitions
now becomes elliptical. The latter is given in polar coordinates by
\begin{eqnarray}
r_{\phi}(\phi)=  \left[ \frac{\hbar  v_{\phi} \Delta/2}{ 4 g_m \mu_B \nabla {\cal B}}\frac{[(\Delta/2)^2 +  U_{e0}^2]} {[(\Delta/2)^2 + U_{e0}^2 \cos^2 \phi]^{3/2}} \right]^{1/2}. \label{elipsoid}
 \end{eqnarray}
Note that this reduces to Eq.~(\ref{quadrum}) in the absence of an
electric field. Substitution of $\phi=0,\pi/2$ in the previous expression allows for the calculation of the semi-axes of the ellipse (i.e., along the $X$- and $Y$-axes). The circular zone of nonadiabatic transitions shrinks along the external electric field and 
extends in the perpendicular direction. 

Assuming that the $Z$-axis (also being perpendicular to the $X$-axis) radius of death is analogous to
 $r^Y_{\phi}$ (except for a scaling factor due to the doubled
gradient), we can estimate the scaling of the total losses 
as a function of the electric field magnitude.  From the previous expressions, we see that losses occur within a completely asymmetric ellipsoidal zone.  The flux of molecules through this ellipsoid per time unit 
constitutes an estimate of the loss rate.  For low values of the electric field magnitude 
$U_{e0}\ll(\Delta/2)$,
the eccentricity of the ellipse in the $XY$ plane 
is given by  $\epsilon =(3/2)^{1/2}(U_{e0}/(\Delta/2))$. If we approximate the ellipsoid and the molecular cloud with spheres~\cite{Cornell95}, then losses due to the transition $|4\rangle \rightarrow |3\rangle$
would be of the same order of magnitude as in the pure magnetic quadrupole trap, with corrections on the order of  $(U_{e0}/(\Delta/2))^2$. 


In conclusion, for electric field magnitudes satisfying both $U_{e0}\ll(\Delta/2)$ and
Eq.~(\ref{electrico1}), the addition of a constant electric field to the trap
does not increase the rate of losses due to nonadiabatic transitions beyond that expected from a magnetic quadrupole trap.

\subsection{Electric quadrupole field plus constant magnetic field}

The final crossed-field configuration we will examine is that of a electric quadrupole field combined with a constant magnetic field.  We examine the losses in this situation in the same manner described above.

\subsubsection{Trap potential}

We again consider an electric quadrupole field
configuration with field distribution $\mbox{\boldmath$\cal E$}
(\bm{r})=\nabla {\cal E}(x,y,-2z)$, where the gradient $\nabla {\cal E}$ is constant.  In the
absence of a magnetic field, there are only two (doubly degenerate) 
distinct surfaces,
one weak-field-seeking and one strong-field-seeking, and the $\Lambda$-doublet prevents a  
connection between the surfaces in the center of the trap.  They can
be visualized by rotating the two energy curves in Fig.~\ref{ZeeS}(b). The primary effect of adding
 a small constant magnetic field, $\mbox{\boldmath$\cal B$}_0 =
(|\mbox{\boldmath$\cal B$}_0|,0,0)$, is to break the degeneracy.  Higher magnetic
fields can modify the connections between different potential energy surfaces.  For example, tuning the magnetic field can bring 
regions of surfaces closer to degeneracy, which can induce linear Stark dependence along
 the  directions perpendicular to $\mbox{\boldmath$\cal B$}_0$, as discussed in Section~\ref{antes},
 or induce conical intersections similar to those in Fig.~\ref{constantel}.

The surfaces in Fig.~\ref{constantma} correspond to our model OH molecule and are generated for fields of $\nabla {\cal E}=5\times10^3$ V/cm$^2$, $|\mbox{\boldmath$\cal B$}_0|=10^3$ G.
  This magnetic field is small enough that no crossings occur between the middle two
surfaces, and the two trapping surfaces are harmonic. The
upper surface is characterized by the following spring constants: 
$k_x=(g_e \mu_e \nabla {\cal E})^2/ (\Delta/2)$, $k_y=(g_e \mu_e \nabla {\cal E})^2 / (\Delta/2+U_{m0})$, and $k_z=(2g_e \mu_e \nabla {\cal E})^2 / (\Delta/2+U_{m0})$, where $U_{m0}=g_m \mu_B |\mbox{\boldmath$\cal B$}_0|$ (the subindex ``0" is added to indicate a constant magnetic field). 

\subsubsection{Losses}

Exact expressions can be found for the probabilities of nonadiabatic
transitions out of the trapping state, $|4\rangle$.  The adiabaticity criterion for
transitions $|4\rangle \rightarrow |1\rangle$ is:\begin{widetext}
\begin{equation}
g_e \mu_e \nabla {\cal E}
 \left[v_{\phi}  \cos \phi  
\left(U_{m0}+\frac{(\Delta/2)^2 + U_e^2}{\sqrt{(\Delta/2)^2 + U_e^2 \cos^2 \phi} }\right)
+ v_r \sin \phi  \left(U_{m0}+\frac{(\Delta/2)^2 }{\sqrt{(\Delta/2)^2 + U_e^2  \cos^2 \phi}}\right)
  \right]  \ll 4  \frac{E_4^{3}}{ \hbar}. \label{exactaconsEB}
\end{equation}\end{widetext}
 For a general $\phi$, which is the angle that the electric
 field makes with the constant magnetic field (assumed to lie 
along the $X$-axis),
this relation converges to the analogous expressions
for the pure electric trap (see Eq.~(~\ref{u1}) and (\ref{u2}))
when $U_{m0}$ goes to zero (we do not present the details of the necessary transformations here, complicated by a location-dependent change of basis). 

 In the case of the azimuthal term in Eq.~(\ref{exactaconsEB}) being alone,  the ratio of the energy gap to the transition probability, thus the adiabaticity, would increase at every point with the addition of a magnetic field, but the contribution of the radial term seems to depend
on the distance.
  The analysis for points very close to the center of the  trap ($U_{e} \ll \Delta/2$)  shows
 that the ratio between the left- and right-hand sides of Eq.~(\ref{exactaconsEB})
scales as $(1+2U_{m0}/\Delta)^{-2}$. As a result, the nonadiabatic transition is actually hindered by increasing the magnetic field. 

Expressions analogous to Eq.~(\ref{exactaconsEB}) can be obtained
for the other two transitions ($|4\rangle \rightarrow |3\rangle$ and $|4\rangle \rightarrow |2\rangle$), but the lengthy analysis will not be presented here.   We therefore  apply Brillouin-Wigner perturbation theory again to consider the electric quadrupole field as
a perturbation to the homogeneous magnetic field near the trap center.  This perturbation theory shows that the transition $|4\rangle \rightarrow |2\rangle$ depends on both the radial and azimuthal velocities. 
  Close to the trap center, both the transition probability and the gap between the potential energy surfaces do not change noticeably due to the addition of the constant magnetic
 field. Dependence with the latter appears only in  second-order terms in the perturbation, $U_e$, far from the center. Thus losses due to this transition are not expected to be very different than the ones in the pure electric quadrupole trap.  However, numerical estimates using the exact expressions show that the presence of the magnetic field
 may inhibit the corresponding spin-flips.

Finally, we consider the transition
 $|4\rangle \rightarrow |3\rangle$. The magnetic field breaks the degeneracy 
in the upper subspace and nonzero transition probabilities due to both radial and azimuthal motion connect both states. Given the confining character of state $ |3\rangle$ they  do not lead to trap losses.
Regardless, while the transition probabilities (which connected already the degenerate states in  the pure electric quadrupole for the  analogous choice of basis) remains essentially constant,
the gap increases linearly
 with the addition of the magnetic field. Thus these flips are again hindered by the presence of the additional
 field.

 In conclusion, the addition of a constant magnetic field to an
 electric quadrupole decreases trap losses beyond those found in a pure electric quadrupole trap. 
 In the previous analysis we have assumed tiered surfaces like that shown in Fig.~\ref{constantma}, where $U_{m0}< \Delta/2$ (magnetic field less than $\sim$1.5 kG for our model OH molecule).  Degeneracies and crossings between surfaces $|2\rangle$ and $|3\rangle$ that occur at higher magnetic fields might deserve special attention.

\section{Remarks on Stark deceleration of OH}

The Stark deceleration technique, an experimental scheme to obtain
samples of cold polar molecules~\cite{Meijer99}, relies also on the absence of significant
nonadiabatic transitions when switching from one electric field stage to the
next.  Molecules transferred to the wrong states would not
experience the proper transverse focusing and slowing and the molecular packet would disintegrate.  We can apply the formalism  above with only minor modifications
to analyze the Stark deceleration process, and in particular, its application to OH in the $^2\Pi_{3/2}$ state~\cite{Bochinski03}.

As the molecule moves along the decelerator, it experiences brisk changes in the local
field. As a result, the field orientation $\alpha$ and strength
 $|\mbox{\boldmath$\cal E$}|$ may change, giving rise to non-vanishing derivatives $d \alpha/ dt$ and $d U_e /
dt$.  
  Eqs.~(\ref{u1}) and (\ref{u2}) can be  modified to include them in our model:
\begin{eqnarray}
|\langle \phi_{t1}|\frac{d}{dt}|\phi_{u1}\rangle|&=&|\langle \phi_{t2}|\frac{d}{dt}|\phi_{u2}\rangle|  =  \label{u11} \\
& &\frac{(\Delta/2)  g_e \mu_e}{2[(\Delta/2)^2 + U_e^2]} \frac{d |\mbox{\boldmath$\cal E$}|}{d t} 
 \nonumber\\
|\langle \phi_{t1}|\frac{d}{dt}|\phi_{u2}\rangle|&=&|\langle \phi_{t2}|\frac{d}{dt}|\phi_{u1}\rangle|  =   \label{u22} \\
& & \frac{U_e}{2\sqrt{(\Delta/2)^2 + U_e^2} } \frac{d \alpha }{d t}  
. \nonumber
\end{eqnarray}

The changes in the local field are maximum when the voltage is
switched between pairs of electrodes.  We estimate
these based on the Stark decelerator and electrostatic trap of the
Lewandowski group at JILA~\cite{Heather}.  During a switching time of
$\sim$1 $\mu$s, the modulus of the local field at the midpoint
of a pair of electrodes decreases a factor of 30 while rotating by
$\pi /2$.  Assuming that the molecule is essentially motionless
during this process, this yields maximum values of $d \alpha /dt = 10^6$ rad/s 
and $d |\mbox{\boldmath$\cal E$}| /dt= 10^{12} $ Vcm$^{-1}$s$^{-1}$.
   Introducing in  Eq.~(\ref{salto2}) the derivatives given by
   Eq.~(\ref{u11}) and (\ref{u22}), we find nonadiabatic probabilities of the order of $10^{-8}$
for OH.  For NH in the $^1 \Delta$ metastable state, another Stark decelerated molecule of interest~\cite{Meijer07b,Heather},  the contribution due to the change of modulus (calculated using Eq.~(\ref{u22})) is several orders of magnitude smaller.  Thus, nonadiabatic transitions
are negligible in the decelerator, a conclusion in agreement with
the findings of \cite{Meijer07b}.

\section{Conclusions}

  The presence of both permanent magnetic and electric dipoles in case (a) molecules opens the way to rich trap dynamics, intrinsically different than that more commonly found in atomic systems.
 The experimental exploration of their behaviour and the potential 
control of their dynamics, which could be exploited through the use 
of arbitrary superpositions of 
electric and magnetic fields, is a question of great importance to current experimental research.
 In particular, we have shown how the simultaneous presence of both fields
can lead to new spatial distributions within the trap. Also, 
while homogeneous electric fields open new nonadiabatic connections  
between states in a magnetic quadrupole trap, constant magnetic
 fields can inhibit losses in an electric quadrupole trap. 
In summary, for practical design of magneto-electrostatic traps, such as those experimentally explored in Ref.~\cite{Sawyer07}, it should be noted that crossed electric and magnetic fields effect
can be used to inhibit previously existing nonadiabatic
transitions by increasing the energy gaps between surfaces or to enhance those transitions by introducing new depolarization pathways.
  The suitable design of the field parameters in magneto-electrostatic traps can reduce nonadiabatic transition loss rates below other limiting loss mechanisms, and thereby open new avenues for explorations of ultracold molecular physics without the need for building unwieldy Ioffe-Pritchard traps.  

 The comparative analysis of nonadiabatic transitions in pure magnetic and electric distributions has also allowed the identification of a new type of nonadiabatic transition in changing electric fields, foreign to the usual
Majorana loss mechanism. Namely, nonadiabatic transition due to the change of magnitude of the electric field and not due to its orientation, and this effect is intrinsically linked to the  presence of the avoided crossing induced by the $\Lambda$-doubling.

We also remark that Hund's case (b) molecules are unaffected by the new nonadiabatic loss mechanisms that appear in Hund's case (a) molecules, and therefore may be more attractive for certain applications.

\acknowledgments
ML and JLB gratefully acknowledge support from the NSF and the W.
M. Keck Foundation, and BLL from the NRC.
We thank J. Ye, H. Lewandowski, B. Sawyer, E. Hudson, E. Meyer and S. Y. T. van de Meerakker for helpful discussions. 
\appendix

\section{Stark and Zeeman interactions in crossed fields}\label{matrix elements}
  When external $\mbox{\boldmath$\cal E$}$-  and $\mbox{\boldmath$\cal B$}$-fields do not form a shared quantization axis, the scalar products which define
 Zeeman and Stark interaction terms have to be evaluated for general $\mbox{\boldmath$\cal E$}({\cal E}_x,{\cal E}_y,{\cal E}_z)$
 electric and  $\mbox{\boldmath$\cal B$}({\cal B}_x,{\cal B}_y,{\cal B}_z)$ magnetic vectors. We outline how matrix elements of the Hamiltonian in Eq.~(\ref{Ham}) may be calculated in these cases.  Both electric and magnetic dipole moments are naturally expressed in the molecular frame, whose $Z$-axis coincides with the molecular
axis.  The electric dipole moment lies along the molecular axis while the magnetic one is easily
 calculated using electronic operators that are referenced to the molecular frame.
Fields, on the other hand, are best expressed in the laboratory frame. The
evaluation of the scalar products which define Stark and Zeeman interaction thus requires the expression of dipoles and fields in the same reference
 system, which we can choose as the laboratory reference system. Changing the spherical components of the dipole
 moments from the molecular to the laboratory frame involves the use of Wigner matrix elements
 ($\mu_q=\sum_k D^{1*}_{qk} \mu_k $). 
 
 Using these notions, the Zeeman and Stark
 interaction terms can be expressed as:
\begin{eqnarray}
H_i=-\bm{\mu} \cdot \mbox{\boldmath$\cal F$}= - \sum_q  (-1)^q (\sum_k
D^{1*}_{qk} \mu_k)  {\cal F}_{-q},\label{inte}
\end{eqnarray}
where $\bm{\mu}$ and $\mbox{\boldmath$\cal B$}$ represent the dipole and the
field, respectively. As the electric dipole moment lies along the
internuclear axis, its spherical coordinates in the
 molecular frame can be expressed as $\bm{\mu}_e=(0,0,\mu_e)$.  If the contributions
of the out-of-axis components of the magnetic dipole moment are
negligible, as we have assumed in the text, 
$\bm{\mu}_m \approx (0,0,-\mu_B (L_0+g S_0)/\hbar)$ is also
obtained.
 Thus, the interaction term~\ref{inte} reduces to
\begin{eqnarray}
H_i= - \sum_q  (-1)^q D^{1*}_{q0} \mu_{0} {\cal F}_{-q}.
\end{eqnarray}
Matrix elements for the interaction operators can be
evaluated using the following expression, in the basis without parity:
\begin{widetext}
 \begin{eqnarray}
 & \langle  J M \Omega |  D^{1*}_{q0}  | J' M' \Omega'  \rangle =
 & \sqrt{2J+1}\sqrt{2J'+1} \left(\begin{array}{ccc} J & 1 &
 J' \\ -M & q & M'\end{array}\right)
\left(\begin{array}{ccc} J & 1 & J'  \\ -\Omega & 0 & \Omega'
\end{array} \right)  (-1)^{M- \Omega}, 
\end{eqnarray}

which,  after some algebra, allows one to obtain the following elements
in the parity defined basis:

\begin{eqnarray}
  \langle  J M \bar{\Omega} \epsilon|  D^{1*}_{q0} \mu_e | J' M' \bar{\Omega'} \epsilon' \rangle = 
 & \left(  \frac{1-\epsilon \epsilon'(-1)^{J+J'+2 \bar{\Omega}}}{2}\right) \sqrt{2J+1}\sqrt{2J'+1} \left(\begin{array}{ccc} J & 1 &
 J' \\ -M & q & M'\end{array}\right)
\left(\begin{array}{ccc} J & 1 & J'  \\ -\Omega & 0 & \Omega'
\end{array} \right)  (-1)^{M- \Omega} \mu_e
\end{eqnarray}
and
\begin{eqnarray}
\langle  J M \bar{\Omega} \epsilon|  D^{1*}_{q0} \mu_B (L_0+g_e S_0)/\hbar| J' M' \bar{\Omega'} \epsilon' \rangle =
\end{eqnarray}

\begin{eqnarray}
  \left(  \frac{1+\epsilon \epsilon'(-1)^{J+J'+2 \bar{\Omega}}}{2}\right) \sqrt{2J+1}\sqrt{2J'+1} \left(\begin{array}{ccc} J & 1 &
 J' \\ -M & q & M'\end{array}\right)
\left(\begin{array}{ccc} J & 1 & J'  \\ -\Omega & 0 & \Omega'
\end{array} \right)  (-1)^{M- \Omega} \mu_B (\Lambda+g_e \Sigma)/\hbar  \nonumber,
\end{eqnarray}
\end{widetext}
where the relation $\bar{\Omega}=\Lambda+\Sigma$ allows one to choose
the correct signs of $\lambda$ and $\sigma$ components in the previous
expression. The difference in the parity factor in both expressions is
related to the magnetic dipole moment not being a constant
(as is $\mu_e$), but an operator. The evaluation of the
 latter---$L_0+g_e S_0$---in the parity basis extracts different signs when acting on the different signed
 $\Omega$ components.

\section{Semiclassical analog for Hund's case (a) molecules}

The competition between the electric and magnetic effects that
determines the potential energy surfaces can be illustrated by a simple analogy that provides useful intuition.  The
electric dipole moment lies parallel to the molecular
axis.  For a case (a) molecule, the magnetic dipole is also
constrained to lie along this axis, on average, due to strong
spin-orbit coupling.  Each field will thus try to orient
the molecule along itself.  

It is reasonable to imagine that the system will essentially 
align to the stronger field.  That being the case, when rotating the smaller field by varying $\alpha$
and holding the magnitudes of both fields fixed, 
the molecular axis will remain anchored to the stronger field while the interaction with the smaller field would vary as the projection of the dipole on the smaller field, $\cos \alpha $.  In fact, dependences of the form $U_m \pm U_e  \cos \alpha $ and $U_e \pm U_m  \cos \alpha $ 
can be extracted from Eq.~(\ref{eigenvalues}) in the limit of
 very small $\Delta$.

\begin{figure}[h]
\begin{center}
\epsfig{file=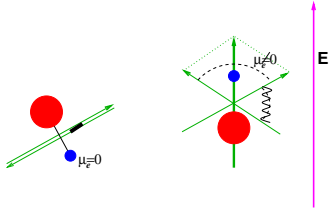,angle=0,width=0.5\linewidth,clip=}
\caption{(Color online) ``Scissors'' model for the polarizability dependence
of the electric dipole  with the electric field magnitude. The molecule
works as a pair of identical dipoles joined by a ficitious spring that
acts to keep
them antiparallel.  As
the field is turned on, both electric dipoles tend to align with the field
but are partially counterbalanced by the spring. The result is a
variable electric dipole moment
whose value depends on the magnitude of the field.} \label{scissors}
\end{center}
\end{figure}
A semiclassical model can also be found for the case in which both fields exert a similar force on the molecule.
As illustrated in  the righthand drawing of Figure~\ref{angulo}, if the
molecular axis makes an angle $\gamma_1$ with respect to $\mbox{\boldmath$\cal E$}$, then
it makes an angle $\alpha-\gamma_1$ with respect to $\mbox{\boldmath$\cal B$}$, and the
resulting energy as a function of orientation is
\begin{eqnarray}
U(\gamma_1)_{\pm}=U_e \cos(\gamma_1)\pm  U_m \cos(\alpha-\gamma_1) \nonumber. \end{eqnarray}
Minimizing (maximizing) this energy with respect to $\gamma_1$ provides the energies given in Eq.~(\ref{eigenvalues}) in the absence of $\Lambda$-doubling. The $\pm$ acounts for the relative orientation of
the electric and magnetic dipoles. 

To include the effect of $\Lambda$-doubling, we imagine that 
the molecule consists of two, instead of one, 
antiparallel electric dipoles
 in zero
electric field.  This corresponds to the fact
that the molecule in zero field experiences no net dipole moment. As
the field is turned on, both electric dipoles tend to align with the field,
but are counterbalanced by a fictitious spring that acts to keep
them antiparallel (see Fig.~\ref{scissors}).  The spring serves to reproduce the $\Lambda$-doublet in this
model.  If the angle between the two electric dipoles is denoted
$\gamma_2$, then this ``scissors'' model can be described by a classical energy
\begin{align}
U_{\pm}(\gamma_1,\gamma_2)=U_e \sin(\gamma_2/2) \cos(\gamma_1) \pm  U_m \cos(\alpha-\gamma_1) & \nonumber  \\
 - \Delta/2
\cos(\gamma_2/2), & 
\end{align} 
where we have introduced particular dependences on $\gamma_2$  
of both the molecule electric dipole moment
 and the energy of the imaginary spring.  The extrema of this function
 constitute a good approximation
to  the quantum energies except that
they cannot reproduce the avoided crossing.


\end{document}